\documentstyle[aaspp,11pt]{article}
\input{psfig}

\begin{document}

\title{RED GIANT--DISK ENCOUNTERS: FOOD FOR QUASARS?}

\author{P.J. Armitage\altaffilmark{1,2}}
\author{W.H. Zurek\altaffilmark{2}}
\author{M.B. Davies\altaffilmark{1}}

\altaffiltext{1}{Institute of Astronomy, Madingley Road, Cambridge,
		CB3 0HA, UK}
\altaffiltext{2}{Theoretical Astrophysics, T-6, MS B288, LANL, 
		Los Alamos, NM 87545, USA}

\begin{abstract}
We explore the role that red giants might play in the fuelling and
evolution of Active Galactic Nuclei. Due to their large radii and
the low binding energy of the stellar envelope, giants are vulnerable
to envelope stripping from collisions with the accretion disk.
Using hydrodynamic simulations we show that such collisions will
typically deposit a substantial fraction of the envelope mass into
the disk on each passage. Repeated encounters will then lead to
the complete destruction of the star save for the dense core. We 
estimate the rate of fuel supply by this mechanism using simple 
models for the AGN disk and central stellar cluster. If the central 
stellar density is $\sim 10^7 \ M_\odot \ {\rm pc}^{-3}$, then 
stripping of giants could account for the activity of typical AGN 
provided that the accretion disk extends out to $\sim 0.1$ pc. 
For AGN with smaller disks, or clusters of lower central density,
giant stripping could supply gas enriched via stellar nucleosynthesis
to a disk replenished from some other source. We find that, for
typical parameters, this mechanism is able to supply important 
quantities of gas to the disk at lower stellar densities than 
previously proposed stellar fuelling models for AGN.

\end{abstract}

\newpage

\section{INTRODUCTION}
Understanding the origin and evolution of the fuel supply that feeds
quasars and other Active Galactic Nuclei (AGN) is a key mystery in
the astrophysics of these systems. To account for luminosities as
high as $\sim 10^{46}$ erg/s the most efficient conceivable 
process - accretion onto a black hole - requires a mass supply
of order one solar mass per year. That matter (which over the
lifetime of the active phase may amount to $10^8 \ M_\odot$ or 
more) must be transported from galactic scales down
to the Schwarzschild radius of the hole at just a few A.U. Evidently
this transport mandates an efficient mechanism for disposing of
the inflowing material's angular momentum.

A partial solution to the problem is provided by invoking the
presence of an accretion disk in the immediate vicinity of
the hole. Although the details are uncertain, viscosity in the
disk transports angular momentum outward, and thereby permits mass 
to flow inward. Recent observations by {\it ASCA} of 
relativistically--broadened iron $K\alpha$ lines lend strong support 
to the idea of the existence of both massive black holes and disks in 
AGN (Tanaka et al. 1995).
For the purpose of fuelling the hole, however, the disk is
unlikely to provide more than a short term reservoir of
mass, as models suggest that the disk mass is only a small
fraction (perhaps $10^{-3}$ for a $10^8 M_\odot$ black hole) 
of the mass of the hole (e.g. Clarke 1988). Even allowing for 
generous uncertainties in our models of accretion disks, therefore, 
it seems inevitable that the disk itself must be continually 
replenished from some further, external, source of mass.

Two possibilities suggest themselves for the source of this
external fuel supply. In the first, the disk could be fed
by gas funnelled in from large (kpc) galactic scales, perhaps
as a consequence of interactions between the AGN host galaxy
and its neighbours (Hernquist \& Mihos 1995). This scenario is
attractive insomuch as there is undoubtably sufficient
mass in gas at those large scales, though whether it can
be transported to the sub-parsec scales of the accretion
disk efficiently and without forming stars in the process
is less clear. As a second possibility, the disk could be
resupplied as a consequence of interactions occurring
in a dense nuclear star cluster. Such clusters are observed
in many well-studied nearby galaxies, with the core
of M32 exhibiting a central stellar density exceeding
$10^7 M_\odot {\rm pc^{-3}}$ (Lauer et al 1992; Crane et
al 1993). Clusters of similar or modestly greater 
($\sim 10^8 \ M_\odot {\rm pc^{-3}}$) richness
then have the potential to be a significant
source of AGN fuel--{\em if} a mechanism exists that
allows the AGN to tap into that supply. 

Disruption of stars in the tidal field of the black hole was
suggested by Hills (1975) as a possible way to liberate the gas.
Disruption occurs when the pericentre $r_{\rm min}$ of the orbit 
becomes comparable or smaller than the `tidal' radius
\begin{equation}
 R_{\rm T} \simeq 5 \times 10^{12} \left({M \over {10^6 M_\odot}}\right)^{1/3} 
 \left({R_* \over R_\odot}\right)
 \left({M_* \over M_\odot}\right)^{-1/3} {\rm cm}
\label{1}
\end{equation}
where $M$ is the hole mass (Rees 1988).
An attractive prediction of this scenario is that activity should
turn off when the central black hole has grown to the point
that $R_{\rm T}$ is comparable to the Schwarzschild radius,
\begin{equation}
 R_{\rm BH} = 2GM/c^2,
\label{2}
\end{equation}
after which stars are swallowed whole without significant release
of radiation. For solar type stars this occurs when the black hole
has reached a mass of a few $\times 10^8 \ M_\odot$. 

For the purpose of long-term fuelling of the black hole, however,
this mechanism proves inadequate. Rapid disruption of stars
passing within $R_{\rm T}$ depletes a region in phase space
which is refilled via relaxation of the stars in the central
cluster on a timescale $\sim 10^8$ yr. This rate is insufficient
to provide enough fuel for the most luminous AGN (Shields
\& Wheeler 1978).

Other possibilities that have been assayed include mass loss 
from stellar winds, which might be enhanced as a result of
the AGN's ultraviolet or high energy particle flux. A generic
problem with such ideas is that gas released
throughout the volume of the nuclear star cluster is vulnerable
to being expelled by the radiation pressure from the central
source. If only a fraction of the liberated mass reaches the
disk or black hole then the demands placed on the supply of mass
are appropriately aggravated.

The interaction between a nuclear star cluster and the accretion flow
has been explored for the case of main sequence stars by a number of
authors. For such stars collisions with the disk will tend
to expel matter out of the disk plane while leaving the star relatively
unscathed. This mechanism has been proposed as an origin for the
gas producing the broad emission lines seen in AGN spectra (Zurek,
Siemiginowska \& Colgate 1991, 1994; see addendum in ApJ XXXX). 
Repeated collisions will gradually
dissipate the star's orbital energy and in time drag it bodily
into the plane of the disk (Syer, Clarke \& Rees 1991). The
consequences of stars captured in this fashion, for example for the
metallicity of quasar accretion disks, have been extensively
investigated (Artymowicz 1993; Artymowicz, Lin \& Wampler 1993).

Here, we extend previous studies on star-disk interactions to
consider the case of giant stars, with radii typically $\sim 100 R_\odot$.
The greatly increased geometric area and reduced binding energy
suggest that such stars should be susceptible to large-scale
stripping of mass on collision with the disk (Zurek, Siemiginowska \&
Colgate 1994). Within some critical radius $R_{\rm crit}$ collisions
will then destroy the star and unbind the envelope mass
within the red giant lifetime. The value of $R_{\rm crit}$ will
depend on the extent and properties of the disk as well as the
velocity and structure of the star, but as we show later it is
invariably larger than the radius within which stars are trapped
while still on the main sequence. If $R_{\rm crit}$ and the central
stellar density are sufficiently large, this mechanism could then
provide an important source of fuel for the central black hole.

The plan of this paper is as follows. In section 2 we discuss
the numerical methods and initial conditions of our calculations. In 
section 3 we present results of hydrodynamic simulations of the star-disk 
interaction for a range of stellar velocities $v_*$, and disk surface 
densities $\Sigma$. Our aim is to delineate that portion of the 
$v_* - \Sigma$ parameter space in which collisions lead to significant 
envelope loss. In section 4 we then combine the results of our calculations
with simple models for the accretion disk and stellar cluster, in
order to estimate how much fuel could be provided to the AGN
by this mechanism. Section 5 summarises our results and conclusions.

\section{NUMERICAL METHOD AND INITIAL CONDITIONS}
The calculations described here were performed using a three-dimensional
smoothed particle hydrodynamics code (SPH), minimally modified from
that described by Benz (1990). This code has been extensively used
in other simulations of stellar collisions (eg Davies, Benz \& Hills
1991), and has been shown to give the same results on such problems
as codes based on alternative numeric techniques (eg PPM--Davies 
et al. 1993). 

In constructing our model star, we commence with a red giant structure
calculated using the latest version of the Eggleton stellar
evolution code (Pols et al. 1995 and references therein). The
model is computed for a mass of $1 M_\odot$, with a core mass
of $0.45 M_\odot$ and a radius $R_* \simeq 150 R_\odot$. We
take this model as being reasonably representative of red giant
structures in galactic nuclei, while acknowledging that a more
complete investigation than is attempted here should consider a 
range of stellar masses and radii. The model also ignores the
possible consequences of the star's proximity to the AGN, though
we note at this stage that the main effect of irradiation on
giants is to {\em increase} their radii (Tout et al. 1989), which
would tend to accentuate envelope loss on collision with
the disk.

\begin{figure}[tb] 
\vspace{0.05truein}
\hspace{1.5truein}
\psfig{figure=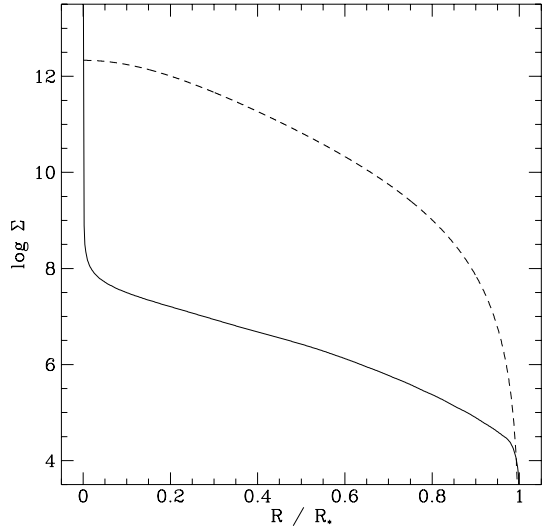,width=3truein,height=3truein}
\caption{Log of the column density $\Sigma$ (g/cm$^2$) as a function
of the fractional stellar radius $R / R_*$. The solid line shows the
profile for the red giant model used in the simulations; the dashed
line represents the profile for an $n=3$ polytrope of equal mass.}
\end{figure} 

The structure of the giant model is shown in Fig. 1, which
plots the column density through the star ($\Sigma_* = \int \rho_* 
{\rm d}z$) as a function of the fractional radius $R / R_*$. For
comparison we also show the structure of a main sequence star of
the same mass, modelled as an $n=3$ polytrope. 
For the red giant the typical column densities near the outer edge
of the envelope are comparable to those seen in models of AGN
accretion disks ($\Sigma \sim 10^4 - 10^5 \ {\rm gcm^{-2}}$), while
for the main sequence star the densities are several orders of
magnitude greater. Thus if each collision strips off an annulus
of mass with column density below some critical threshold (and
we show later that this is indeed a reasonable estimate), then
the red giant is vastly more susceptible to mass loss. This
is quantified in Fig. 2, which plots the mass exterior
to a given column density for the two models. At the relatively
low $\Sigma$ values found in AGN disks, the giant has a 
vulnerable mass fraction more than four orders of magnitude
in excess of a main sequence star of the same mass. This implies
that a star may survive repeated disk transits on the main
sequence yet be destroyed relatively swiftly on reaching the
red giant phase of its evolution.

\begin{figure}[tb] 
\vspace{0.05truein}
\hspace{1.5truein}
\psfig{figure=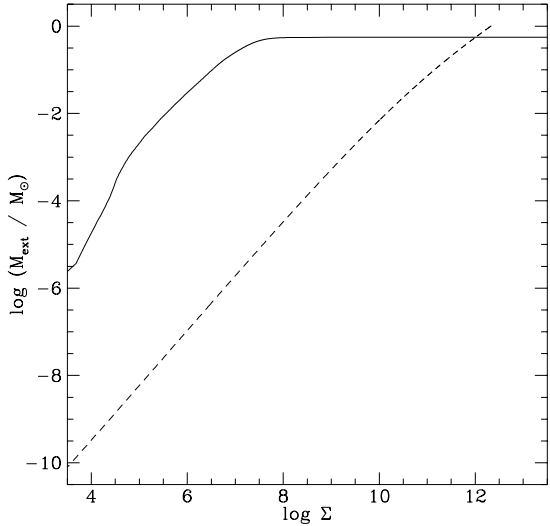,width=3truein,height=3truein}
\caption{Integrated stellar mass exterior to a cylindrical radius
with given column density $\Sigma$. The solid line shows the
profile for the red giant model used in the simulations; the dashed
line represents the profile for an $n=3$ polytrope of equal mass.}
\end{figure} 

\begin{figure}[tb] 
\vspace{0.05truein}
\hspace{1.5truein}
\psfig{figure=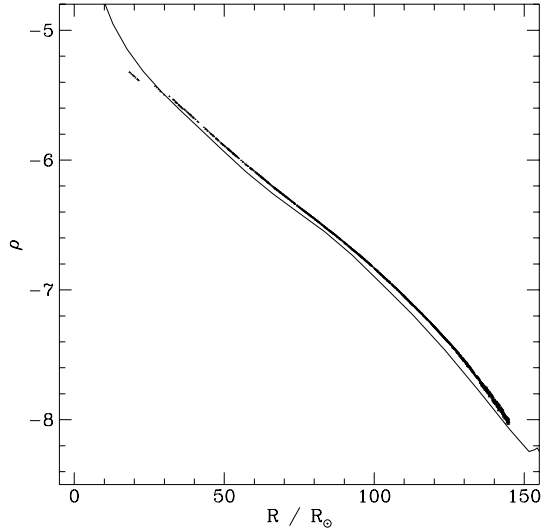,width=3truein,height=3truein}
\caption{Log of the stellar density $\rho / {\rm gcm^{-3}}$ 
for the red giant used in our simulations as a
function of radius. Solid line is the output of the stellar
model, the symbols depict $\rho$ at the positions of the particles
in the SPH realisation.}
\end{figure} 

To represent the stellar structure in SPH, we replace the core
of the giant with a single point mass that interacts with the
remaining particles via gravitational forces only. The envelope
is modelled using 2274 SPH particles that are set down on a
close-packed grid with masses proportional to the local
density in the model star. As the density in an SPH calculation
is not a local quantity this is a crude procedure, but it
gives a density run that matches the stellar model to about
20\% over 2 orders of magnitude in density (Fig. 3), which is 
is adequate for our purposes here. The internal energy 
and pressure of the particles are then set to ensure 
hydrostatic equilibrium, using an equation of state that
includes a mixture of gas and radiation pressure,
\begin{equation}
 P = {{\rho {\cal R} T} \over \mu} + {{\alpha T^4} \over 3},
\label{3}
\end{equation}
where $\mu = 0.6$ is the mean molecular weight and $\alpha$ the
radiation constant. We have verified that stars modelled using
this procedure are stable for at least several sound crossing times,
which is an order of magnitude longer than the duration of the
star-disk simulations.

For the disk we adopt a simple approach. The timescale for the
star to transit the disk is only $\sim 10^6$ s, which is negligible
compared to any timescale in the disk. We therefore represent
the disk as a column of initially cold gas, with a gaussian density
profile in the vertical direction (corresponding to an isothermal
temperature distribution). Disk models suggest that the thickness
of the disk is at least an order of magnitude greater than the
size of even our giant star, so we pick a scale height $H$ such that
the density gradient across the star is small. For our main
simulations we take $H = 600 R_\odot$, and truncate the disk 
in the vertical direction at 1.5 scale heights. We also experiment
with other choices for $H$. With these parameters, we achieve
comparable resolution in the star and disk with around 19,000
disk particles. 

\section{RESULTS}

Table 1 lists the main parameters for the SPH simulations of
star-disk encounters, the stellar velocity $v_*$, disk surface 
density $\Sigma$, and scale height $H$. We also tabulate the
impacting disk mass $M_i = \pi R_*^2 \Sigma$, the incident
momentum $p_i$, and kinetic energy $E_i$. As the binding
energy of the model stellar envelope is only $1.6 \times
10^{46}$ erg, {\em all} of these collisions have the potential 
in principle to unbind a large fraction of the stellar material.
The final column in the table shows the main result of these
calculations--the mass loss from the red giant--evaluated at
a time 0.5 disk transit times after the star exits the disk.
For all except the most violent collisions (which have anyway
almost totally destroyed the stellar envelope by this stage) the
induced mass loss has ceased by this time.

Figure 4 ({\em omitted from this draft})
shows a montage of frames from simulation E,
where at each snapshot we have plotted particles within 2 SPH
smoothing lengths of the $y=0$ plane. This is the most energetic
encounter, and as such demonstrates most clearly the two effects
leading to mass loss. Firstly, momentum transfer between the
disk gas and the star can bodily strip material from the sides
of the star. This reduces the cross-sectional area the star 
presents to the disk, and creates a wake composed primarily
of stellar material behind the star. Second, the disk gas 
impinging on the front surface of the star shock heats and
compresses the material there. When the star emerges from the 
disk the heated gas then blows off as an almost spherical
expanding cloud. These general hydrodynamic properties
of the collision are very similar to those seen in simulations
of stars impacted by shells of supernova ejecta (Fryxell \&
Arnett 1981; Taam \& Fryxell 1984; Livne, Tuchman \& Wheeler
1992). For supernovae that explode with red giant companions,
indeed, the energy and momentum of the encounter are similar
to those considered here. The main difference is then the
relative thickness of the disk compared to the star--unlike
shells of supernova ejecta typical AGN disks are always many stellar
radii thick.

\subsection{Mass loss}

\begin{figure}
\caption{Omitted from this draft due to size}
\end{figure}

\begin{figure}[tb] 
\vspace{0.05truein}
\hspace{1.5truein}
\psfig{figure=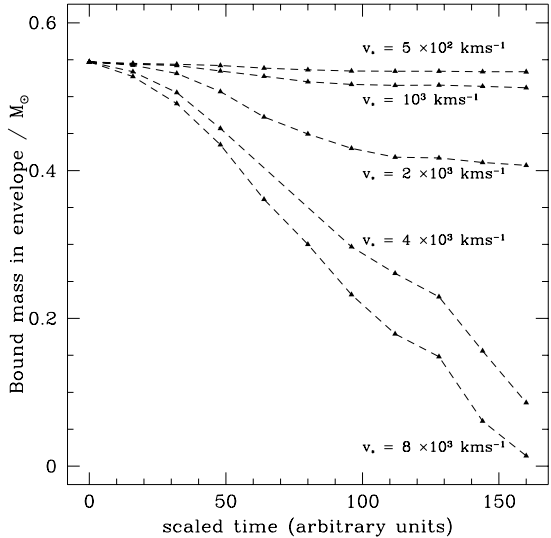,width=3truein,height=3truein}
\caption{Bound mass in the stellar envelope as the star passes through
the disk is shown here for runs with different impact velocities. The 
time axis has been scaled so that for each simulation the disk transit
time is constant at 100 units. From top to bottom, the impact velocities
were $v_* =$ 500 km/s, 1000 km/s, 2000 km/s, 4000 km/s and 8000 km/s. 
For each case the disk surface density was $10^5 \ \rm{g cm^{-2}}$.}
\end{figure} 

Figure 5 shows for runs A-E the bound mass in the stellar 
envelope as the star crosses the disk. We compute the bound mass
at regular intervals by summing the masses of particles whose
total energy (gravitational plus internal and kinetic) is less
than zero. These simulations all have
the same disk surface density ($10^5 \ {\rm gcm^{-2}}$) and scale
height ($H = 600 R_\odot$), and differ only in the stellar
velocity. To facilitate comparison between the runs, we scale
the time axis so that the disk crossing takes place between
(in arbitrary time units) $t=10$ and $t=110$ for all the simulations.

From the figure, it can be seen that the behaviour divides into
two regimes. The two most energetic collisions (D and E) disrupt
the stellar structure entirely. Mass loss during the disk
crossing is rapid, and most of the gas in the remnant that emerges
from the disk is itself formally unbound from the white dwarf core.
This prompt destruction of the star as a consequence of a single
disk passage occurs at this disk surface density for velocities
of around 4000 km/s or greater. For AGN, that implies radii in the 
disk of less than 6000 $R_{BH}$, which as we shall show in Section
4.2 encompasses all the non-self-gravitating disk in a supermassive
system of $10^8 \ M_\odot$.

Less violent impacts (A-C) produce smaller fractions of stripped
and ablated mass, and leave the inner regions of the star relatively
unaffected. In these cases the rate of mass loss declines rapidly
once the star has exited the disc, and so the final bound mass
in our simulations is likely to give a good estimate of the total
mass loss from the disk encounter. We find mass loss that ranges
from $0.14 \ M_\odot$ for the $v_* = 2000$ km/s run down to
approximately 1\% of a solar mass for the $v_* = 500$ km/s
simulation.

\subsection{Comparison with analytic predictions}

To further interpret our simulations, we compare our numerical results
to the semi-analytic theory developed by Wheeler, Lecar \& McKee (1975,
henceforth WLM) for the analogous situation of a star impacted by
a supernova blast wave. Although we have already noted that there
are important differences between that scenario and the star-disk
encounter, the WLM theory has the merit of providing a simple
baseline against which to compare results. In the case of
supernovae the theory has been shown to agree closely with detailed
calculations (Fryxell \& Arnett 1981).

Following WLM, we identify two components to the total mass loss.
First, direct momentum transfer strips matter from the outside of
the star down to some fraction of the stellar radius $x_{crit}$.
This critical radius is that where the momentum transfered to a
cylindrical shell is just sufficient to accelerate it to the stellar 
escape velocity $v_{\rm es}$ at that radius. As the impacts we
are considering are highly supersonic ($v_* \gg v_{\rm es}$) this
condition implies that mass is lost exterior to the cylinder where
\begin{equation}
 \Sigma_*(R) = \left( v_* \over {v_{\rm es}(R)} \right) \Sigma.
\label{3.15}
\end{equation}
The quantity $\Sigma_*(R)$ is the column density through the star as
plotted in Fig. 1. The stripped mass fraction 
by this direct mechanism, $F_{\rm strip}$, is then the total mass in the
envelope exterior to a cylinder of radius $R_* x_{crit}$. From our
red giant model $x_{\rm crit}$ and hence $F_{\rm strip}$ are
straightforward to evaluate for a given incident momentum $p_i$.

Calculating the second, ablation component of the mass loss requires
computing the heating generated by the shock that is driven into the
star as a consequence of the collision. As the star exits the ejecta 
shell or disk, this thermal energy will become available as kinetic
energy as the heated material `boils off' the stellar surface. WLM
give the result,
\begin{equation}
 F_{\rm ablate} \approx \Psi x_{\rm crit}^2,
\label{3.1}
\end{equation}
where their parameter $\Psi$ is given in our notation by,
\begin{equation}
 \Psi = M_i \left( {v_* \over v_{\rm es}} - 1 \right).
\label{3.2}
\end{equation}
For highly supersonic encounters, $\Psi$ is thus a measure of the
incident momentum. 

\begin{figure}[tb] 
\vspace{0.05truein}
\hspace{1.5truein}
\psfig{figure=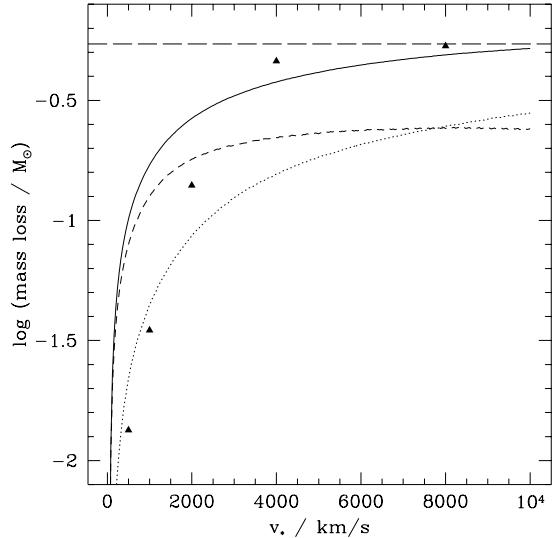,width=3truein,height=3truein}
\caption{Stellar envelope mass loss as a function of encounter velocity.
The symbols show the results of SPH simulations, the lines represent
the expected mass loss using the WLM approximation (see text). Dotted
line is $F_{\rm strip}$, dashed line is $F_{\rm ablated}$, and the solid
line the total mass loss predicted. The dashed horizontal line represents
total loss of the stellar envelope.}
\end{figure} 

Figure 6 compares the WLM predictions with our
numerically derived mass loss estimates for runs A--E. At high 
velocities of 4000 km/s or greater there is reasonable agreement 
between the mass loss seen in our simulations and the total
predicted by the WLM analysis $F_{\rm total} = F_{\rm strip} + 
F_{\rm ablate}$. Both agree that at these velocities essentially all
of the stellar envelope becomes unbound as a consequence of the
collision.

At lower impact energies and momenta, we find that $F_{\rm strip}$
alone provides a good estimate of the mass loss seen in the simulations,
suggesting that for these lower velocity ($v_* \leq 1000 \ km/s$) 
encounters mass loss is driven primarily by momentum transfer 
rather than ablation. This transition as the velocity decreases
is plausible on physical grounds--the large thickness of the disk
relative to the stellar radius means that the collision is less
like a single impulsive impact and more closely resembles a steady
flow, albeit one in which the star undergoes steady stripping. Thus
compared to the supernova case the strength of shock heating and
ablation is liable to be reduced, leaving the directly stripped
component dominant at low velocity. In support of this
interpretation we note that repeating run C with the disk scale
height halved (run I) leads to a doubling of the mass loss. Whereas
run C shows mass loss intermediate between $F_{\rm strip}$ and
$F_{\rm total}$, run I exhibits mass loss given closely by
$F_{\rm total}$. Table 2 shows the results for all the runs
described here.

From our simulations we can then identify two regimes. For high
velocities and/or small disk scale heights the mass loss is large
and can be approximated well by the total mass loss predicted by
WLM, $F_{\rm total}$. This result agrees with the supernova studies
referred to above. Conversely for lower velocity encounters with
disks that are thick compared to the stellar size the mass loss
is substantially reduced below the WLM prediction. In this case 
we can take $F_{\rm strip}$ alone as a reasonable estimate of the
mass loss that will occur in the collision.

\subsection{Post-encounter stellar structure}

For the purpose of supplying fuel to the disk, the most important 
collisions are those that occur at relatively low velocity at a 
large radius in the disk. Although the mass loss per collision 
is small, the cumulative effect of many such impacts dwarfs that 
of the violent impacts that unbind the envelope in a single disk
transit. We quantify this statement in the following Section, but
note here that it is only true if the equilibrium structure of
the perturbed star is {\em at least} as vulnerable to stripping
as the original model. If this is the case we can regard the
mass loss predicted for a given $v_*$ and $\Sigma$ by our
simulations as a {\em lower} limit on the mass loss in subsequent
disk encounters.

To calculate the post-encounter stellar structure we first note that
radiation pressure is negligible in the bound stellar envelope of the
remnant, so that the polytropic exponent $\gamma$ is everywhere close 
to $5/3$. From our SPH simulations we can then extract the polytropic
constant $K$ at the location of each particle,
\begin{equation}
 K = { P \over {\rho^{\gamma}} }.
\label{3.3.1}
\end{equation}
Following the impact $K$ displays considerable scatter as a function
of radius. This is a result of the non-symmetric injection of entropy 
from the shock-heating of the envelope. We therefore average over
angles to obtain a smooth approximation to $K(R)$ for use in 
calculating the structure. We also compute the mass interior to 
radius $R$ to express $K$ as a function of enclosed mass $m$.
Physically we expect the star to settle
to a new equilibrium with a spherically symmetric $K$ distribution
on of order the convective timescale--much shorter than the
interval between collisions with the disk. Provided that this is the
case our averaging procedure and assumption of spherical symmetry 
should be valid.

Combining the equation representing hydrostatic equilibrium,
\begin{equation}
 { {{\rm d}P} \over {{\rm d}R} } = - { {Gm\rho} \over {R^2} },
\label{3.3.2}
\end{equation}
with equation (\ref{3.3.1}) above then gives an expression for the
stable density gradient in terms of the function $K(m)$,
\begin{equation}
 { {{\rm d}\rho} \over {{\rm d}R} } = 
 - { {Gm\rho^{2-\gamma}} \over {\gamma K(m) R^2} }.
\label{3.3.3}
\end{equation}
This equation together with the known total mass of the remnant
constitute an eigenvalue problem for the central density, which
we solve using a shooting method to yield the equilibrium density
profile.

\begin{figure}[tb] 
\vspace{0.05truein}
\hspace{1.5truein}
\psfig{figure=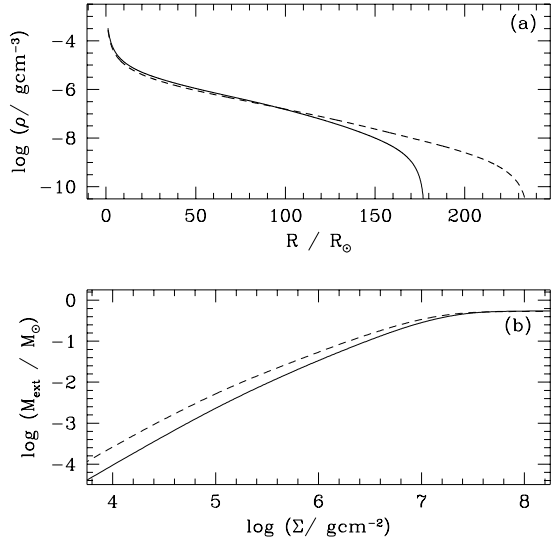,width=3truein,height=3truein}
\caption{Equilibrium stellar structure calculated from the entropy
profile of the red giant before (solid line) and after (dashed line)
the encounter with the disk (run A). The upper panel shows the density
profile, the lower panel the amount of mass exterior to a given column 
density as in Fig. 2.}
\end{figure} 

The results of this analysis are shown in Fig. 7(a) for
run A--the lowest velocity encounter considered here. The density
profile of the inner regions ($R < 100 \ R_\odot$) is predicted
to be almost unaltered by the impact, as expected since the
influence of the collision on the star is confined to the
outer layers for the less violent impacts. The main effect
of the increased entropy generated by the encounter is to
puff--up the outer envelope, leading to an increase in
the radius of about 30\%.

Fig. 7 illustrates how the recomputed stellar
model differs from the initial red giant in its vulnerable
mass fraction, expressed as the mass $M_{\rm ext}$ exterior
to a cylinder of column density $\Sigma$ (cf Fig. 2).
The post-encounter star is {\em more} vulnerable to stripping
than the initial model, by factors of around 2 for the typical
values of column density considered here. We 
are therefore confident that repeated star--disk encounters will act
cumulatively to disrupt the stellar envelope, and regard the
estimates given in Section 4 (where we assume that the total mass
loss is linear in the number of impacts) as conservative predictions
of the rate at which destruction occurs.   
 
\section{IMPLICATIONS FOR THE FUELLING OF AGN}
In the previous section we computed the mass stripping associated
with a variety of star-disk collision parameters. The most violent
collisions ($v_* \geq 4000$ km/s, $\Sigma = 10^5 \ {\rm gcm^{-2}}$)
lead to essentially complete stripping of the stellar envelope
on a single disk transit. For less violent encounters the mass loss
and disruption of the stellar structure is greatly reduced. In
this regime we have used our results to validate a simple estimation
procedure for the mass loss, based on the amount of momentum imparted
to the stellar envelope by the impact. 

In this Section we utilise the above results to investigate the 
significance of this mechanism for the fuelling of AGN. In Section 4.1 
we calculate the radius out to which a disk could strip stars of their
envelopes within the giant lifetime--this determines the volume of
the cluster that is vulnerable to disk stripping. We compare this
to the radii at which the competing processes of tidal disruptions; 
star-star collisions; and stellar trapping would operate. 
In Section 4.2 we calculate thin disk models to determine the 
`target area' of the disk which has surface density sufficient 
to cause stripping. Finally in Section 4.3 we estimate the
total mass that this mechanism might provide to fuel AGN
activity for a variety of central stellar densities, and
use the results of stellar evolution calculations to predict
the evolution of the mass supply with time.

\subsection{Determining the critical radii in the disk}

On each disk transit, the star will be stripped down to a column
density $\Sigma_*$ given by equation (\ref{3.15}). The corresponding
mass loss can be directly read off from Fig. 2, which is
well approximated for small $M_{\rm loss}$ by
\begin{equation}
 M_{\rm loss} \sim { v_* \over v_{\rm es} } \left( \Sigma \over
 10^4 {\rm gcm^{-2}} \right) \times 10^{-5} \ M_\odot.
\label{4.2.1}
\end{equation} 
The timescale between encounters for stars orbiting at a radius $R$ is 
\begin{equation}
 t_{\rm orbit} \approx \pi \sqrt{ R^3 \over {GM} },
\label{4.2.2}
\end{equation}
where $M = M_{\rm BH} + M_{\rm c}$ is the total enclosed mass at that 
radius including the contribution from the cluster itself. Imposing the
condition that at $R_{\rm crit}$ the star will just be stripped during
the red giant lifetime $t_{\rm rg}$ then yields,
\begin{equation}
 R_{\rm crit} \sim 0.05 \ {\rm pc} \times \left( M \over {10^6 M_\odot}
 \right)^{1/2} \left( t_{\rm rg} \over {10^7 {\rm yr}} \right)^{1/2}
 \left( \Sigma \over {10^4 {\rm gcm^{-2}}} \right)^{1/2}
 \left( v_{\rm es} \over {50 {\rm kms^{-1}}} \right)^{-1/2},
\label{4.2.3}
\end{equation}
where we have inserted reasonable values for all the parameters. A
$10^8 \ M_\odot$ black hole is therefore capable of stripping red
giants out to a distance of several tenths of a parsec, {\em if} the
disk matter itself extends to that radius.

The red giant stripping process discussed here will evidently be
important only if stars survive to become giants without first
being destroyed by other processes, such as tidal disruption,
stellar trapping, or star--star collisions. The radius at which
stars are disrupted by the tidal field of the black hole is 
given by equation (\ref{1}); we now estimate the critical 
radii for the other two competing processes.

In evaluating the timescale for stars to be ground into the 
plane of the disk we follow closely the approach of Syer, Clarke
\& Rees (1991). If the only mechanism leading to decay of the
orbit is direct hydrodynamic drag on passing through the disk,
then to order of magnitude the trapping timescale will be the 
time required for the star to impact its own mass of disk material. 
If on each passage the star collides with a mass $\Delta m = \pi
R_*^2 \Sigma$,
\begin{equation}
 t_{\rm trap} = \chi { m_* \over {\Delta m} } t_{\rm orbit},
\label{4.2.4}
\end{equation} 
where $\chi \sim 5$ is a dimensionless scaling factor that depends
on the initial orbital parameters (Syer, Clarke \& Rees 1991). Stars
will be brought into the disk plane provided that $t_{\rm trap}$ is
shorter than the relaxation time $t_{\rm relax}$, after which stars 
will typically have diffused onto less radial orbits. This criteria
gives a radius interior to which stars will be trapped,
\begin{equation}
 R_{\rm trap} \sim 10^{-3} \ {\rm pc} \times \left( M \over {10^6 M_\odot}
 \right)^{1/3} \left( t_{\rm relax} \over {10^8 {\rm yr}} \right)^{2/3}
 \left( R_* \over R_\odot \right)^{4/3} \left( \Sigma \over {10^4 
 {\rm gcm^{-2}}} \right)^{2/3} \left( m_* \over M_\odot \right)^{-2/3},
\label{4.2.5}
\end{equation}
which is typically intermediate between the tidal disruption and 
giant stripping radii.

For the dense stellar systems we are considering, physical collisions
may destroy stars before they even reach the giant stage of their
evolution. We note at this stage that physical collisions might themselves 
be important in fuelling AGN, however they are undoubtably a different type
of mechanism to that considered here and require detailed 
study to determine, for example, the efficiency of mass loss.
For stars of a single representative radius $R_*$, a simple
expression for the collision rate $\Gamma$ is,
\begin{equation}
 \Gamma = \pi R_*^2 v_* n(R),
\label{4.2.6}
\end{equation}  
where $n(R)$ gives the run of the stellar number density with radius. 
For simplicity, we assume a stellar profile that is a power-law
in radius, with a stellar number density $n(R) = n_0 R^q$. For
an isothermal sphere $q = -2$, while more detailed cusp solutions
for stellar clusters containing a central black hole have $q = -7/4$
(Binney \& Tremaine 1987 and references therein). A condition for 
collisions to be important at a radius $R$ is that 
$\Gamma (R) \tau_{MS} \simeq 1$--in which case 
stars within that radius will be destroyed
before completing their main sequence life $\tau_{MS}$. For 
$q = -2$ this radius is
\begin{equation}
 R_{\rm col} \sim 10^{-2} \ {\rm pc} \times \left( \tau_{MS} \over
 {10^8 {\rm yr}} \right)^{2/5} \left( \overline{m} \over M_\odot
 \right)^{-2/5} \left( M_c \over {10^7 M_\odot} \right)^{2/5}
 \left( M \over {10^6 M_\odot} \right)^{1/5} \left( R_* \over 
 R_\odot \right)^{4/5},
\label{4.2.7}
\end{equation}
where $\overline{m}$ is the mean stellar mass, $M_c$ is the cluster 
mass enclosed within a radius of 1 pc, and $M$ is the total mass 
within radius $R$. 

\begin{figure}[tb] 
\vspace{0.05truein}
\hspace{1.5truein}
\psfig{figure=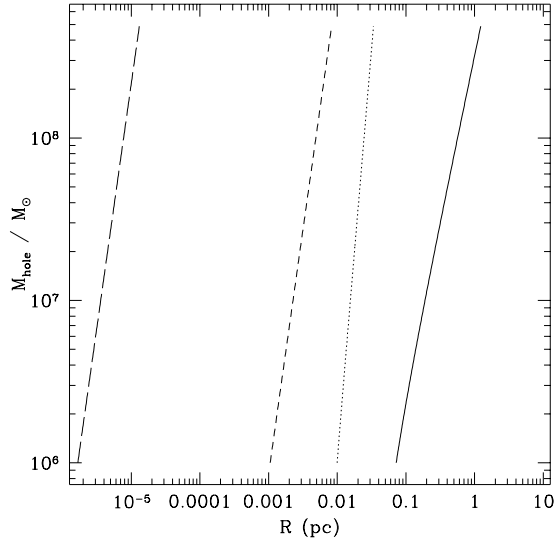,width=3truein,height=3truein}
\caption{Radii at which various stellar fueling mechanisms become
important for different masses of black hole. From left to right;
tidal disruption (long dashes), stellar trapping (short dashes),
star--star collisions (dots) and giant stripping (solid). This
plot shows radii calculated for a cluster with $M_c = 10^7 \ 
M_\odot$, with a disk of $\Sigma = 10^4 \ {\rm gcm^{-2}}$ extending
to large radius.}
\end{figure} 

Fig. 8 shows these radii as a function of black hole mass.
For this plot we have taken parameters $t_{\rm rg} = 10^7$ yr,
$t_{\rm relax} = 10^8$ yr, and $\tau_{MS} = 10^8$ yr. We assume
a cluster with a mass $M_c = 10^7 \ M_\odot$ within 1 pc, and
a disk with a surface density $\Sigma$ of $10^4 \ {\rm gcm^{-2}}$
(ignoring for now the complication of having $\Sigma$ itself a
function of radius). There is a large parameter space here, but
from the figure it is evident that the red giant stripping
radius $R_{\rm crit}$ exceeds the other radii plotted by at
least an order of magnitude for all black hole masses in the
illustrated range. This implies, in particular, that most
stars that are vulnerable to stripping as giants
{\em will} survive to reach the red giant phase
without suffering prior destruction by physical collisions. 
We note here that our estimate for the collision time
on the main sequence is consistent with that of Begelman
\& Sikora (1991), who concluded that models in which 
red giants generated Broad Emission Lines from their 
stellar winds could be ruled out, because the stars would
collide before reaching the giant phase. The apparent
difference with our result is caused by the requirement for such models,
(in which the stellar line-producing envelopes must
provide a covering fraction towards the central source
of $\sim 0.1$), to invoke a vastly greater stellar
density in red giants than is required here.

Experimentation with the simple models presented above shows that, for most
values of the parameters, we obtain plots qualitatively similar
to Fig. 8, except that for lower central stellar
densities the collision radius is reduced and lies interior
to the trapping radius. We therefore conclude that for a disk
that extends to large radius stellar stripping of giants will
provide a greater source of fuel to the disk than the other
stellar mechanisms considered here. 

\subsection{Determining the target area of the disk}

To determine the typical extent and properties of AGN accretion disks, 
we need the run of surface 
density with radius in the disk. To calculate this, we follow the
approach of Clarke (1988) to construct vertically averaged thin
disk solutions, which give the radial dependence of the surface
density, sound speed, and other disk quantities. Although it is
possible to go beyond vertically averaged models by solving
for the details of the disk vertical structure, such calculations
are more `accurate' only if the site of energy dissipation in
the disk is known. As this in turn depends on the unknown 
processes generating the viscosity, we cannot in practice
expect to do better than the simple vertically averaged
models presented here. Our models use the `alpha-prescription'
for the viscosity (Shakura \& Sunyaev 1973); for completeness
we set out the thin disk equation solved here in an Appendix.

\begin{figure}[tb] 
\vspace{0.05truein}
\hspace{1truein}
\psfig{figure=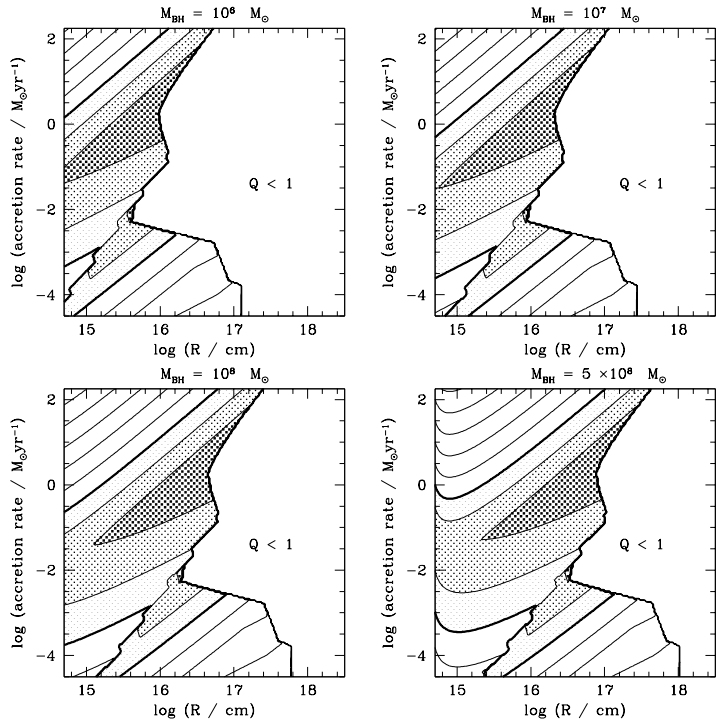,width=4truein,height=4truein}
\caption{Contour plot of the disk surface density as a function of
radius $R$ and accretion rate through the disk $\dot{M}$. The panels
depict different masses for the central black hole from $10^6 \ M_\odot$
(top left), up to $5 \times 10^8 \ M_\odot$ (bottom right).
Contours are plotted at intervals of 0.5 in $\log(\Sigma / {\rm gcm^{-2}})$,
with the highest surface density regions shaded.
The bold contour is at $\Sigma = 10^4 {\rm gcm^{-2}}$. The uncontoured
region to the right of each plot is where the Toomre parameter, $Q$, is
less than unity.}
\end{figure} 

Figure 9 shows contour plots of the disk surface density 
$\Sigma$ as a function of radius and mass transfer rate $\dot{M}$ 
in the disk, for black hole masses that cover the range generally 
considered for AGN. These models assume constant $\dot{M}$ with
radius, and an $\alpha$ of 0.1, similar to the inferred values
in dwarf nova disk systems. The lower ($\sim 10^{-2}$) values
favoured by some authors from analyses of UV variability
(Siemiginowska \& Czerny 1989) would increase the surface density
in the disk, and thus somewhat enhance the efficacy of the 
stripping process. At each radius in our solutions, we also
evaluate the value of the Toomre $Q$ parameter, which measures
the importance of self-gravity,
\begin{equation}
 Q = {{c_s \Omega} \over {\pi G \Sigma}}.
\label{4.6}
\end{equation}
In this equation $c_s$ is the mid-plane sound speed and $\Omega$
the Keplerian angular velocity in the disk. If $Q < 1$ the disk
becomes unstable to local gravitational instabilities, and
although the structure in this regime is uncertain, the disk
may develop spiral density waves or break up into clumps. As 
in such a situation the angular momentum is {\em not} transported
via an $\alpha$ viscosity, consistency requires that we truncate
our solutions when $Q$ falls below unity.

Fig. 9 shows that for accretion rates between 
$10^{-3}$ and $10^1$ $M_\odot \ {\rm yr^{-1}}$ the surface
density in the disk will be of order $10^4 \ {\rm gcm^{-2}}$ or
greater. This is the case for all the black hole masses considered
here. However we also find that the disk would be expected to
become self-gravitating at relatively small radii, ranging from
$\sim 10^{16}$ cm ($3 \times 10^{-3}$ pc) for the $10^6 \ M_\odot$
hole to around $\sim 10^{17}$ cm for the $10^8 \ M_\odot$ case. The
properties of the disk beyond these radii are unclear, though
calculations by Laughlin \& Bodenheimer (1994) for self-gravitating
protostellar disks suggest that gravitational torques transport
angular momentum with an efficiency equivalent to an $\alpha$ of
around 0.03. If this applies also to AGN disks, then it suggests
that the average $\Sigma$ in the self-gravitating regime should
be at least as large as that calculated from the Shakura-Sunyaev
analysis above.

\subsection{Fuelling history}

From Fig. 8 we find that supermassive black holes should
be able to strip mass from red giants out to radii of $\sim 0.1$ pc
provided only that their accretion disks extend to that radius. A
0.1 pc disk with average $\Sigma = 10^4 \ {\rm gcm^{-2}}$ has a
mass of $\sim 10^6 \ M_\odot$, which is still only a small fraction
of the hole mass for the larger mass black holes. If the cluster
mass $M_c = 10^7 \ M_\odot$, then the mass of stars intersecting
the disk will be of order $10^6 \ M_\odot$, with the precise number
dependent on the distribution of stellar orbits in the cluster
(with more radial orbits leading to a greater fraction of vulnerable
stars). If this mass of stars on orbits susceptible to stripping
is fixed, then the rate at which this mass is liberated to the disk is
controlled purely by stellar evolutionary processes. 
Once a given star reaches the red giant phase of its evolution
its envelope mass will rapidly be deposited into the disk.

To investigate the evolution predicted by this model, 
we shall assume that the stellar population in the nuclear star
cluster can be described as a coeval population with a power-law
Initial Mass Function (IMF)
\begin{equation}
 N(m) \propto m^{-\nu}. 
\label{4.2}
\end{equation}
Typically we take $\nu = 1.35$, corresponding to a Salpeter IMF, and
impose a lower mass cut-off at the hydrogen-burning limit ($m \simeq
0.1 M_\odot$). If after a time $\tau$ (where $\tau = 0$ is the time
of the initial star formation) stars of mass $m = m(\tau)$ are leaving 
the main sequence and becoming giants, then the rate at which mass is being
deposited into the disk is
\begin{equation}
 \dot{M} = {{{\rm d}N} \over {{\rm d}m}} {{{\rm d}m} \over {{\rm d}\tau}}
 \left( m(\tau) - m_{\rm core} \right).
\label{4.3}
\end{equation}
In this expression $m_{\rm core}$ is the core mass which is not captured by
the disk, while the envelope mass has been assumed to be instantaneously
deposited into the disk once the star has evolved off the main sequence.
This should be a reasonable approximation given that the red giant lifetime
is much shorter than that on the main sequence. 

Once the hydrogen burning stellar lifetime $\tau(m)$ is specified, 
equation (\ref{4.3}) gives the mass deposition rate as a function
of time. To gain some insight into the prediction of this model 
for the fuelling rate, we first assume that $\tau(m)$ is a power-law 
in the stellar mass. For masses of a few $M_\odot$, a crude
estimate is,
\begin{equation}
 \tau(m) \simeq 10^{10} \ {\rm yr} (m / M_\odot)^{-3}.
\label{4.4}
\end{equation}
At early times, when $m \gg m_{\rm core}$, the rate of deposition of AGN fuel
is then given by
\begin{equation}
 \dot{M} \sim \tau^{(\nu - 4)/3} \sim \tau^{-0.9},
\label{4.5}
\end{equation}
where the final step uses for definiteness $\nu = 1.35$. The fuelling
history is dominated by the steep dependence of the stellar lifetime
on the mass, so that the mass deposition rate declines roughly
inversely with time. If all other parameters of the system remained
fixed (for example disk radius and surface density), we would then
expect the supply of fuel today to be only a thousandth of the rate
at an early ($10^7$ yr) epoch. Rapid evolution with cosmic time is,
of course, a well-known property of AGN.

\begin{figure}[tb] 
\vspace{0.05truein}
\hspace{1.5truein}
\psfig{figure=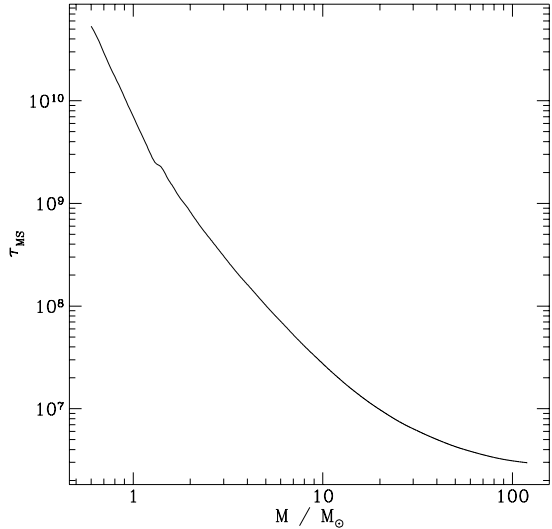,width=3truein,height=3truein}
\caption{Main sequence lifetime as a function of stellar mass, from
calculations by Fagotto et al. (1994).}
\end{figure} 

Figure 10 shows the results of detailed calculations of the
hydrogen burning lifetime $\tau(m)$ by Fagotto et al. (1994), 
for stars in the mass range 0.6 $M_\odot < m <$ 120 $M_\odot$ . 
These calculations assume a low metallicity ($Z = 0.004$) and employ
modern (OPAL) opacities. Combining this function with an assumed
IMF yields curves for the mass deposition rate, and the fraction of
total mass liberated
by a given time, that are shown in Fig. 11. Here we take 
$m_{\rm core} = 0.45 \ M_\odot$, and vary the cluster mass
and slope of the IMF.

\begin{figure}[tb] 
\vspace{0.05truein}
\hspace{1.5truein}
\psfig{figure=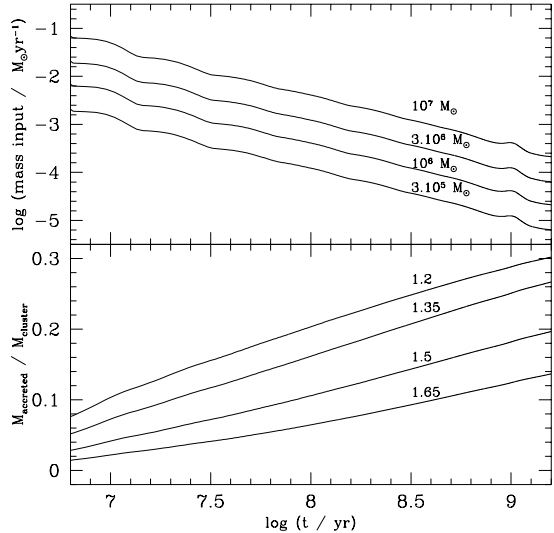,width=3truein,height=3truein}
\caption{Upper panel: rate of mass deposition in the disk as a function of
time assuming a constant disk surface density profile. The curves
are plotted for different masses of stars on vulnerable orbits that
intersect the disk, from top downwards $10^7 \ M_\odot$,
$3 \times 10^6 \ M_\odot$, $10^6 \ M_\odot$ and $3 \times 10^5 \ M_\odot$.
The lower panel shows the fraction of the stellar mass that has been
stripped by a given time for various assumed IMFs, from top downwards
we assume $\nu = $ 1.2, 1.35, 1.5 and 1.65 respectively.}
\end{figure} 

For a cluster mass that intersects the disk of $10^7 \ M_\odot$, we
obtain a deposition rate at $10^7$ yr of nearly $10^{-1} \ M_\odot
{\rm yr^{-1}}$, which declines by approximately 2 orders of magnitude
by $10^9$ yr. For IMFs that are
close to the Salpeter value, we find that about 20-30\% of the
initial cluster mass has been released to the disk within that time, 
though this is fairly sensitive to the assumed IMF. A steep IMF
locks up a large fraction of the total mass in
long-lived low-mass stars, and thus lowers the fraction of
mass that reaches the giant phase and is supplied to the disk.
We also note that 
the effect of an extended period of star formation would be
to lower and `smear out' the period of peak mass deposition. This
would occur if the high mass star formation epoch exceeded 
approximately $10^7$ yr.

Typical AGN luminosities are of the order of $10^{44}$ erg/s, which 
corresponds to an accretion rate of $\sim 10^{-2} \ M_\odot \ {\rm yr}^{-1}$
(assuming a 10\% efficiency of converting rest mass into
radiation). Fig. 11 then shows that mass deposition rates of this
magnitude can be supplied from the early evolution of a cluster 
that has a few $\times 10^6 -
10^7 \ M_\odot$ of stars on vulnerable, disk-intersecting orbits. For the
accretion disk radii of $\sim 0.1$ pc postulated above, this requires
a central stellar density within 1pc of the hole of order $10^7 \ M_\odot$.

An immediate prediction of this model is that the rate of mass
deposition into the disk should evolve rapidly. Fig. 11 is
calculated assuming that the target area of the disk remains
fixed, and so strictly represents the time-evolution of the mass
supply if the disk is primarily replenished from some source
other than star-disk collisions. In such a scenario, the main
effect of the stripping would be to enrich the disk with the
products of stellar nucleosynthesis. Such enrichment would be
most significant at early times, shortly after the cluster
was formed, and would subsequently drop off rapidly as shown in the Figure.
Even faster evolution is predicted if star-disk collisions
{\em dominate} the mass supply to the disk, as in this instance
the radius to which the disk can strip stars would itself
decrease with time. The evolutionary track would then
move toward the lower curves in the Figure, as fewer and
fewer stars became vulnerable to destruction via
disk impacts. Although the details would depend on the (poorly known)
disk physics, the general scenario is then one in which a burst
of star formation in the cluster leads to a short, (perhaps $10^8$
yr, as by then the accretion rate would have fallen to the
values where $\Sigma$ starts dropping rapidly - Fig. 9), 
but intense flare in the AGN activity. As the mass deposition
drops, so the disk weakens until it can no longer strip stars,
and the activity then ceases altogether.

\section{DISCUSSION}
In this paper we have presented simulations of encounters of red
giant stars with AGN accretion disks. We find that for encounters
within around 0.1 pc of the black hole, and disk surface
density exceeding $10^4 \ {\rm gcm^{-2}}$, the fractional mass loss
on collision is sufficiently large that the star would not survive
the red giant phase. Rather, the envelope mass would be deposited
into the disk from the cumulative effect of numerous disk transits,
leaving only a stripped core orbiting the black hole. The most
violent collisions achieve this end in only a single passage. Although
we have considered only one type of giant, we believe these conclusions 
are likely to apply equally to a range of red giant masses and structures.

We have compared the radius out to which red giants can be stripped
via encounters with the disk with the radii at which stellar
collisions and trapping by the disk become important. The stripping
radius is generally one or two orders of magnitude greater than
$R_{\rm col}$ and $R_{\rm trap}$, and indeed is most likely to 
be limited by the actual physical extent of the accretion disk.
Even if the stellar density in the nucleus is relatively low, this
mechanism could provide an important source of gas that has been
enriched as a result of stellar nucleosynthesis. Artymowicz (1993)
has suggested that heavy elements might be generated {\em in situ}
by trapped stars accreting and then exploding as supernovae within
the disk. The larger radius at which stripping can occur means 
that a similar quantity of enriched material could be provided
to the disk by this process. 

An immediate consequence of our models is that stars in the innermost
regions of galaxies harbouring accretion disks will have their
lifetime as red giants severely curtailed by star-disk collisions.
The resultant modification of the stellar population is probably
on too small a scale to be observable, though the depletion of the
red giant population close to the black hole does cast further
doubt on models that seek to invoke giants as important
components in the central regions of AGN. Conversely, the rapid
destruction of giants will {\em enhance} the rate of production
of white dwarfs, which will remain as a dense cluster 
on disk-intersecting orbits.
As suggested in a recent paper by Shields (1996), these single
white dwarfs may accrete sufficient gas on passage through 
the disk to greatly enhance the rate of novae, and this provides
yet a further mechanism for enriching the disk gas. We note that
in principle better observations of the chemical mix in the
centers of AGN could constrain which, if any, of these mechanisms -- giant
stripping, novae or supernovae -- were important in supplying
heavy elements to the gas.

How important might this mechanism be in fulfilling the {\em overall}
fuelling requirements of quasars and other AGN? Since the stellar
densities present in the cores of AGN remain unknown, it is
impossible to answer this question with any great certainty. We
find that typical AGN luminosities could be achieved from this
mechanism if the central stellar densities are as high as 
$10^7 \ M_\odot {\rm pc^{-3}}$ or greater, and the accretion
disk extends to a few tenths of a pc. These are {\em lower}
densities than are required for collisions between stars to 
liberate equivalent quantities of fuel. The activity of AGN
powered by this fuel source would decline rapidly as the mass of stars
becoming giants dropped, and would eventually switch off entirely
as the disk became too tenuous to strip further stars. Such 
a scenario is not inconsistent with the evolutionary properties
of quasars and AGN.

Although we have concentrated here on feeding the disk via a 
single population of stars in a central cluster,
we also note that higher total masses of
fuel could be provided by this mechanism if the depleted phase space
could be replenished--perhaps by stars driven inward from the
host galaxy by some means. This is of course essentially the 
`interaction-driven' model of fuelling that we have previously
mentioned. In such a scenario, the results given here imply
that it is irrelevant whether the mass arrives at the nucleus
in the form of gas or stars--even in the latter case a reasonable
fraction of the total mass will be readily made available to feed the 
AGN activity.

\subsection*{Acknowledgements}
We thank Jose Font-Roda, Chris Fryer and Cathie Clarke for commenting on 
earlier versions
of this work, and Chris Tout for supplying the stellar model. PJA thanks
T--6, LANL, for their hospitality over the summer. MBD acknowledges
support from a Royal Society URF. This work was partially
supported by IGGP, LANL. 

\section*{APPENDIX: THIN DISK EQUATIONS}
In Section 4.2 we have presented solutions to the vertically averaged
thin disk equations. These are set out in, for example, Faulker, Lin
\& Papaloizou (1983), and follow the approach of Shakura \& Sunyaev
(1973). Explicitly we use the following forms. Local thermal equilibrium
of an isolated annulus of the disk implies,
\begin{equation}
 2 \sigma T_e^4 = {9 \over 4} \nu \Sigma \Omega^2,
\label{a1}
\end{equation}
where $T_e$ is the disk effective temperature, $\sigma$ the Stefan-Boltzmann
constant, $\nu$ the kinematic viscosity, and $\Omega$ the Keplerian 
angular velocity. A vertically averaged treatment of the disk radiative
transport gives a relation between the effective temperature and the 
central temperature $T_c$,
\begin{equation}
 T_e^4 = {{8 T_c^4} \over {3 \kappa \Sigma}}
\label{a2}
\end{equation}
where $\kappa$ is the midplane opacity. The local mass flux $\dot{M}$
is
\begin{equation}
 \nu \Sigma = {\dot{M} \over {3 \pi}} \left(1 - \sqrt{R_* / R} \right)
\label{a3}
\end{equation}
for a disk with a zero-torque ($\Sigma = 0$) inner boundary condition
at a radius $R = R_*$. The alpha prescription for the viscosity is
conventionally expressed as,
\begin{equation}
 \nu = {2 \over 3} {{\alpha P} \over {\Omega \rho}}
\label{a4}
\end{equation}
with $\alpha$ a free parameter and $\rho$ the central disk density. In
lieu of greater knowledge of the form of the viscosity in AGN disks,
we take $\alpha$ to be a constant, typically 0.1. The equation of
state is as previously used,
\begin{equation}
 P = {{{\cal R} \rho T_c} \over \mu} + {{4 \sigma} \over {3c}} T_c^4.
\label{a5}
\end{equation}
Finally we have the equation representing hydrostatic equilibrium in the vertical direction,
\begin{equation}
 H = {\Sigma \over {2 \rho}} = {c_s \over \Omega}
\label{a6}
\end{equation}
where $c_s$ is the midplane sound speed and $H$ is the disk scale height.

The set of equations (\ref{a1}) - (\ref{a6}) can be readily solved
once the opacity $\kappa$ is specified. We utilise the analytic fits
given by Bell \& Lin (1994)--these are identical to those used by
Lin \& Papaloizou (1985) except for modifications to the low (below
3000K) temperature regime suggested by Alexander et al. (1989).
Detailed expressions are as given in Bell \& Lin (1994). 

\subsection*{REFERENCES}

\noindent Alexander, D.R., Augason, G.C., \& Johnson, H.R., 1989, 
ApJ, 345, 1014

\noindent Artymowicz, P., 1993, PASP, 105, 691

\noindent Artymowicz, P., Lin, D.N.C., \& Wampler, E.J., 1993, ApJ, 409, 592

\noindent Bell, K.R., \& Lin, D.N.C., 1994, ApJ, 427, 987

\noindent Begelman, M.C., \& Sikora, M., 1991, in AIP Conf.
Proc. 254, Testing the AGN Paradigm, ed. S.S. Holt, S.G. Neff, \& 
C.M. Urry (New York: AIP)

\noindent Benz, W., 1990, in The Numerical Modelling of Nonlinear Stellar
Pulsations, ed. J.R. Buchler, Kluwer Academic Publishers, Dordrecht, p. 269

\noindent Binney, J., \& Tremaine, S., 1987, Galactic Dynamics, Princeton
University Press, Princeton, p. 545

\noindent Clarke, C.J., 1988, MNRAS, 235, 881

\noindent Crane, P., 1993, AJ, 106, 1371

\noindent Davies, M.B., Benz, W., \& Hills, J.G., 1991, ApJ, 381, 449

\noindent Davies, M.B., Ruffert, M., Benz W., \& M\"uller, E., A\&A, 272, 430

\noindent Fagotto, F., Bressan, A., Bertelli, G., \& Chiosi, C., 1994, 
A\&AS, 105, 29

\noindent Faulkner, J., Lin, D.N.C., \& Papaloizou, J., 1983, MNRAS, 205, 359

\noindent Fryxell, B.A., \& Arnett, W.D., 1981, ApJ, 243, 994

\noindent Hernquist, L., \& Mihos, J.C., 1995, ApJ, 448, 41

\noindent Hills, J.G., 1975, Nature, 254, 295

\noindent Lauer, T.R., et al., 1992, AJ, 104, 552

\noindent Laughlin, G., \& Bodenheimer, P., 1994, ApJ, 436, 335

\noindent Lin, D.N.C., \& Papaloizou, J., 1985, in Protostars and Planets II,
eds. D.C. Black \& M.S. Matthews, University of Arizona Press, Tucson, p. 981

\noindent Livne, E., Tuchman, Y., \& Wheeler, J.C., 1992, ApJ, 399, 665

\noindent Pols, O., Tout, C.A., Eggleton, P.P., \& Han, Z.W., MNRAS, 274, 964

\noindent Rees, M.J., 1988, Nature, 333, 523

\noindent Shakura ,N.I., \& Sunyaev, R.A., 1973, A\&A, 24, 337

\noindent Shields, G.A., 1996, ApJ, in press

\noindent Shields, G.A., \& Wheeler, J.C., 1978, ApJ, 222, 667

\noindent Siemiginowska, A., \& Czerny, B., 1989, 239, 289

\noindent Syer, D., Clarke, C.J., \& Rees, M.J., 1991, MNRAS, 250, 505

\noindent Taam, R.E., \& Fryxell, B.A., 1984, ApJ, 279, 166

\noindent Tanaka, Y., et al., 1995, Nature, 375, 659

\noindent Tout, C.A., Eggleton, P.P., Fabian, A.C., \& Pringle, J.E., 
MNRAS, 238, 427

\noindent Wheeler, J.C., Lecar, M., \& McKee, C.F., 1975, ApJ, 200, 145

\noindent Zurek, W.H., Siemiginowska, A., \& Colgate, S.A., 1991, in AIP Conf.
Proc. 254, Testing the AGN Paradigm, ed. S.S. Holt, S.G. Neff, \& 
C.M. Urry (New York: AIP)

\noindent Zurek, W.H., Siemiginowska, A., \& 
Colgate, S.A., 1994, ApJ, 434, 46; addendum ApJ, to appear Oct 10 issue

\begin{table}
\caption{RESULTS OF SPH STAR-DISK SIMULATIONS}
\smallskip
\begin{tabular}{llllllll} \hline \hline
Run & $v_*$ & $\Sigma$   & $H$       & $M_i$ & $p_i$ & $E_i$ &$M_{\rm loss}$\\
    & (km/s)& (g/cm$^2$) & ($R_\odot$) & ($M_\odot$) & (gcm/s) & (erg) &
 ($M_\odot$) \\
\hline

A & 500 & $10^5$ & 600 & $1.9 \times 10^{-2}$ & $1.9 \times 10^{39}$ & 
  $4.7 \times 10^{46}$ & $1.3 \times 10^{-2}$ \\

B & 1000 & $10^5$ & 600 & $1.9 \times 10^{-2}$ & $3.8 \times 10^{39}$ &
  $1.9 \times 10^{47}$ & $3.5 \times 10^{-2}$ \\

C & 2000 & $10^5$ & 600 & $1.9 \times 10^{-2}$ & $7.5 \times 10^{39}$ &
  $7.5 \times 10^{47}$ & $1.4 \times 10^{-1}$ \\

D & 4000 & $10^5$ & 600 & $1.9 \times 10^{-2}$ & $1.5 \times 10^{40}$ &
  $3.0 \times 10^{48}$ & $4.6 \times 10^{-1}$ \\

E & 8000 & $10^5$ & 600 & $1.9 \times 10^{-2}$ & $3.0 \times 10^{40}$ &
  $1.2 \times 10^{49}$ & $5.3 \times 10^{-1}$ \\

F & 1000 & $10^4$ & 600 & $1.9 \times 10^{-3}$ & $3.8 \times 10^{38}$ &
  $1.9 \times 10^{46}$ & $2.9 \times 10^{-3}$ \\

G & 4000 & $10^4$ & 600 & $1.9 \times 10^{-3}$ & $1.5 \times 10^{39}$ &
  $3.0 \times 10^{47}$ & $5.8 \times 10^{-2}$ \\

H & 4000 & $2.5 \times 10^4$ & 600 & $4.7 \times 10^{-3}$ & 
  $3.8 \times 10^{39}$ & $7.5 \times 10^{47}$ &  $1.2 \times 10^{-1}$ \\

I & 2000 & $10^5$ & 300 & $1.9 \times 10^{-2}$ & $7.5 \times 10^{39}$ &
  $7.5 \times 10^{47}$ & $2.6 \times 10^{-1}$ \\

\end{tabular}
\end{table}

\begin{table}
\caption{COMPARISON WITH ANALYTIC ESTIMATES}
\smallskip
\begin{tabular}{lllllll} \hline \hline
Run & $v_*$ & $\Sigma$   & $H$       & $M_{\rm loss}$  & $F_{\rm strip}$
    &$F_{\rm wlm}$ \\
    & (km/s)& (g/cm$^2$) & ($R_\odot$) &($M_\odot$)&($M_\odot$)&($M_\odot$)\\
\hline

A & 500 & $10^5$ & 600 & $1.3 \times 10^{-2}$ &
	$2.2 \times 10^{-2}$ & $1.0 \times 10^{-1}$ \\

B & 1000 & $10^5$ & 600 & $3.5 \times 10^{-2}$ &
	$4.5 \times 10^{-2}$ & $1.7 \times 10^{-1}$ \\

C & 2000 & $10^5$ & 600 & $0.14$ &
	$8.6 \times 10^{-2}$ & $0.27$ \\

D & 4000 & $10^5$ & 600 & $0.46$ &
	$0.16$ & $0.38$ \\

E & 8000 & $10^5$ & 600 & $0.53$ &
	$0.25$ & $0.49$ \\

F & 1000 & $10^4$ & 600 & $2.9 \times 10^{-3}$ &
	$4.3 \times 10^{-3}$ & $3.0 \times 10^{-2}$ \\

G & 4000 & $10^4$ & 600 & $5.8 \times 10^{-2}$ &
	$1.9 \times 10^{-2}$ & $9.1 \times 10^{-2}$ \\

H & 4000 & $2.5 \times 10^4$ & 600 & $0.12$ &
	$4.7 \times 10^{-2}$ & $0.17$ \\

I & 2000 & $10^5$ & 300 & $0.26$ &
	$8.6 \times 10^{-2}$ & $0.27$ \\

\end{tabular}
\end{table}

\end{document}